\documentclass[journal]{IEEEtran}

\usepackage[T1]{fontenc}
\usepackage{amssymb}
\usepackage{amsmath}
\usepackage{algorithm}
\usepackage{algorithmic}
\usepackage{graphicx}
\usepackage{epstopdf}
\DeclareGraphicsExtensions{.eps}
\usepackage{caption}

\usepackage{subfigure}
\usepackage{cite}

\begin{document}
%
\title{Hybrid Precoding for Physical Layer Multicasting}
\author{\IEEEauthorblockN{Mingbo Dai\IEEEauthorrefmark{1}, Bruno Clerckx\IEEEauthorrefmark{1}\IEEEauthorrefmark{2}}
\IEEEauthorblockA{\IEEEauthorrefmark{1}Department of Electrical and Electronic Engineering, Imperial College London, UK\\\IEEEauthorrefmark{2}School of Electrical Engineering, Korea University, Seoul, Korea}}
\author{Mingbo~Dai,~\IEEEmembership{Student Member,~IEEE,} Bruno~Clerckx,~\IEEEmembership{Member,~IEEE,}
\thanks{M. Dai and B. Clerckx are with the Department of Electrical and Electronic Engineering, Imperial College London, UK, SW7 2AZ UK (e-mail: \{m.dai13, b.clerckx\}@imperial.ac.uk). B. Clerckx is also with the School of Electrical Engineering, Korea University, Seoul, Korea.}}

\maketitle

\begin{abstract}
This work investigates the problem of downlink transmit precoding for physical layer multicasting with a limited number of radio-frequency (RF) chains. To tackle the RF hardware constraint, we consider a hybrid precoder that is partitioned into a high-dimensional RF precoder and a low-dimensional baseband precoder. Considering a total transmit power constraint over the RF chains, the goal is to maximize the minimum (max-min) received signal-to-noise ratio (SNR) among all users. We propose a low complexity algorithm to compute the RF precoder that achieves near-optimal max-min performance. Moreover, we derive a simple condition under which the hybrid precoding driven by a limited number of RF chains incurs no loss of optimality with respect to the fully digital precoding case. Finally, numerical results validate the effectiveness of the proposed algorithm and theoretical findings.

\end{abstract}

\begin{IEEEkeywords}
Multicasting, Limited RF chains, Hybrid precoding, Low complexity algorithm.
\end{IEEEkeywords}

\IEEEpeerreviewmaketitle

\section{INTRODUCTION}

Wireless multicasting has been emerging as a key enabling technology to efficiently address the overwhelming traffic demands (e.g., popular video delivery to a number of mobile devices) in the next generation cellular networks. The optimization problem of transmit beamforming for quality-of-service (QoS) requirements and for max-min fairness was proven to be NP-hard \cite{Tom2006}. This NP-hard optimization problem can be approximated by a convex semidefinite programming (SDP) problem using a semidefinite relaxation (SDR) approach \cite{Tom2006}. However, the solution of the relaxed problem is not always feasible for the original problem. To address this issue, several iterative algorithms were proposed in \cite{Gopa2015, Nils2011, Tran2014}. In order to facilitate applications for real-time systems, \cite{Joung2015} proposed a non-iterative and simple-yet-effective linear precoding strategy. Besides, the problem of multicasting was extended to multiple co-channel groups \cite{Tom2008}, per-antenna power constraint \cite{Dim2014, Dim2015} and massive MIMO deployments \cite{Xiang2014}, respectively.

With perfect knowledge of channel state information at the transmitter (CSIT), all the aforementioned works assume digital baseband precoding which requires a dedicated radio-frequency (RF) chain for each antenna element at the base station (BS). Unfortunately, such a requirement is very costly and therefore unrealistic for massive MIMO systems (corresponding to a large number of RF chains) and Millimeter-wave (Mm-wave) MIMO systems (due to expensive Mm-wave mixed-signal components). To address the RF hardware constraint, authors in \cite{Omar2014} proposed a hybrid precoding scheme for a single-user MIMO system where the BS implements the hybrid precoder by a high-dimensional RF precoder using cost-efficient analog phase shifters, cascaded with a low-dimensional digital baseband precoder. Moreover, a very recent work \cite{Choi2015} applied hybrid precoder into physical layer multicasting. They computed the hybrid precoder as a sparse weighted combination of predefined vectors and minimizes the $\ell_1$-norm of the weights, rather than solving the original problem of minimizing the transmission power under the QoS constraints.

This letter studies the max-min fairness of multicasting driven by a limited number of RF chains. Consider a predefined RF codebook, a low complexity search algorithm is proposed to determine the RF precoder which obtains nearly the same performance as performing an exhaustive search. Moreover, by exploiting channel sparsity, we prove a simple condition under which the hybrid precoder with a limited number of RF chains achieves the same max-min fairness as with fully digital precoder. The problem of hybrid precoding for multicasting is considered in both Rayleigh fading channels and limited scattering (e.g., Mm-wave) channels.

\emph{Organization:} Section \ref{system} introduces the system model and formulates the problem. Section \ref{HP} details our proposed approaches. In Sections \ref{numresults} and \ref{conclusion}, we present numerical results and conclusions, respectively.

\emph{Notations:} Bold lowercase (uppercase) letters represent vectors (matrices). The notations {\small$[\mathbf{X}]_{i,j}, \mathbf{X}^T, \mathbf{X}^H,\mathbb{E}(\mathbf{X}),\lambda_{\text{max}}(\mathbf{X})$} denote the entry in the $i$-th row and $j$-th column, transpose, conjugate transpose, expectation and the largest eigenvalue of a matrix {\small$\mathbf{X}$}. We use $\text{Span}(\mathbf{X})$ to denote the column space of $\mathbf{X}$ and $\text{Span}^{\bot}(\mathbf{X})$ for its orthogonal complement. $\|\mathbf{x}\|$ indicates the 2-norm of a vector.

\section{PROBLEM FORMULATION} \label{system}
Consider the downlink of a multiuser cellular system where a BS equipped with $M$ antennas and $N$ ($N \le M$) RF chains serves $K$ single-antenna users. Let {\small$\mathbf{h}_k \in \mathbb{C}^{M }$} denote the frequency-flat quasi-static downlink channel vector of user $k$ and assume that the BS has perfect knowledge of the channel vector for each user. The input-output analytical expression is written as {\small$y_{k} = \mathbf{h}_{k}^H \mathbf{x} + n_k$}, where $\mathbf{x} = \mathbf{w} s$ represents the transmitted signal with power constraint {\small$\mathbb{E}[||\mathbf{x}||^2] \le P$}. {\small$\mathbf{w} \in \mathbb{C}^{M}$} is the linear precoder while $s$ is the common message intended to all users with zero-mean and unit variance. {\small$n_k \sim \mathcal{CN} (0,\sigma_{k}^2)$} denotes the additive white Gaussian noise.

\subsection{A Sufficient Number of RF Chains $(N = M)$} \label{SDR}
In this case, $\mathbf{w}$ can be fully designed in the digital domain. Let us revisit the traditional max-min SNR fairness problem \cite{Tom2006} which can be written as
\begin{eqnarray} \label{eq:fullydig}
\underset{\mathbf{w}}{\text{max}} \;\; \underset{k \in \mathcal{K}}{\text{min}} \quad |\bar{\mathbf{h}}^H_{k} \mathbf{w}|^2, \quad \text{s.t.} \;\; \|\mathbf{w}\|^2 = P,
\end{eqnarray}
where {\small$\bar{\mathbf{h}}_k = \mathbf{h}_k/\sigma_k$} and $\mathcal{K}$ denotes the multicast user set. This NP-hard problem can be approximated in a relaxed or conservative manner. Namely, we can relax problem \eqref{eq:fullydig} to a convex SDP problem by introducing {\small$\mathbf{W} = \mathbf{w} \mathbf{w}^H$} and ignoring the rank one constraint. If {\small$\mathbf{W}_{\text{opt}}$} has rank of one, its principal eigenvector is the optimal solution to problem \eqref{eq:fullydig}. Otherwise, a randomization procedure \cite{Tom2006} is applied to obtain a feasible solution to the original problem. For example, we perform eigen-decomposition on {\small$\mathbf{W}_{\text{opt}} = \mathbf{U} \mathbf{\Sigma} \mathbf{U}^H$} and generate a set of candidate vectors as {\small$\mathbf{w} = \mathbf{U} \mathbf{\Sigma}^{1/2} \mathbf{v}$} with {\small$\mathbf{v} \sim \mathcal{CN} (\mathbf{0},\mathbf{I})$} and normalize them as {\small $\|\mathbf{w}\|^2 = P$}. The best candidate which obtains the largest max-min SNR is chosen as the final solution of problem \eqref{eq:fullydig}. By contrast, we can exploit the fact that the max-min fairness problem subject to a transmit power constraint is equivalent to the transmit power minimization problem subject to QoS constraints up to scaling \cite{Tom2006}. Then, replacing the non-convex (QoS) constraint by a conservative convex approximation yields a feasible solution to the original problem \cite{Nils2011}. Besides, \cite{Joung2015} designed the precoder as a linear sum of channels of all users which achieves near-optimal max-min fairness. The corresponding details can be found in \cite{Nils2011, Tran2014, Joung2015}.


\subsection{A Limited Number of RF chains $(N < M)$} \label{lowcomplex}
In this case, $\mathbf{w}$ can be jointly designed in the analog and digital domain, i.e., {\small$\mathbf{w} = \mathbf{F}_{\text{RF}} \mathbf{w}_{\text{BB}}$} where {\small$\mathbf{F}_{\text{RF}} \in \mathbb{C}^{M \times N}$} and $\mathbf{w}_{\text{BB}} \in \mathbb{C}^{N}$. Since {\small$\mathbf{F}_{\text{RF}}$} is implemented using phase shifting networks, a constant modulus constraint is imposed on its entries. Without loss of generality, we assume that {\small$[\mathbf{F}_{\text{RF}}]_{i,j} = \frac{1}{\sqrt{M}} e^{j\varphi_{i,j}}$}. Since $P$ is immaterial with respect to the optimization problem, we can normalize it and then scale up the solution with {\small $\sqrt{P}$}. Proceeding with the design of {\small$\mathbf{F}_{\text{RF}} \mathbf{w}_{\text{BB}}$}, problem \eqref{eq:fullydig} can be stated as
\begin{eqnarray} \label{eq:hybridpre}
&& \underset{\mathbf{F}_{\text{RF}}, \mathbf{w}_{\text{BB}} }{\text{max}} \;\; \underset{k \in \mathcal{K}}{\text{min}} \quad |\bar{\mathbf{h}}^H_{k} \mathbf{F}_{\text{RF}} \mathbf{w}_{\text{BB}}|^2 \nonumber \\
&& \;\; \text{s.t.} \quad  \|\mathbf{F}_{\text{RF}} \mathbf{w}_{\text{BB}}\|^2 = 1, \quad \mathbf{F}_{\text{RF}} \in \mathcal{F},
\end{eqnarray}
where $\mathcal{F}$ is the feasible set of $\mathbf{F}_{\text{RF}}$ with constant magnitude entries.  The NP-hard problem becomes more challenging in the presence of the non-convex feasibility constraint $\mathbf{F}_{\text{RF}} \in \mathcal{F}$.

\section{PROPOSED SOLUTIONS} \label{HP}
Consider a predefined RF codebook and an optimized $\mathbf{w}_{\text{BB}}$ for a given RF precoder, a low complexity search algorithm is proposed to determine the RF precoder that achieves nearly the same performance as performing an exhaustive search. When it comes to limited scattering channels, the RF precoder can be designed exploiting channel sparsity. We derive a simple condition under which the hybrid precoder driven by a limited number of RF chains achieves the same max-min fairness as with fully digital precoder.

\subsection{A Low Complexity Algorithm} \label{algorithm}
To resolve the non-convex constraint, the RF precoder can be selected from a predefined codebook {\small $\mathcal{C}$}. We denote $\mathcal{C}_{\text{set}}$ as the set of all $M \times N$ matrices whose columns are drawn from $N$ different columns of $\mathcal{C}$. Since the RF precoder belongs to a finite discrete set, the associated combinatorial optimization of $\mathbf{F}_{\text{RF}}$ can be solved by a high-complexity exhaustive search with {\small$I_{\text{exs}} = \binom{N}{M}$} iterations. For each $\mathbf{F}_{\text{RF}}$ and the associated effective channel $\hat{\mathbf{h}}_k \triangleq \mathbf{F}^H_{\text{RF}} \bar{\mathbf{h}}_{k}$, we determine the optimal $\mathbf{w}_{\text{BB}}$ by using methods presented in section \ref{SDR}. Then, the $\mathbf{F}_{\text{RF}}$ and $\mathbf{w}_{\text{BB}}$ that achieve the best max-min SNR performance is adopted as the final solution. To circumvent the issue of complexity, we develop a low complexity search algorithm by leveraging the following observation on the RF precoder design.

\emph{Observation}: Problem \eqref{eq:hybridpre} can be equivalently formulated as
\begin{eqnarray} \label{eq:hybridpre2}
&& \underset{\mathbf{F}_{\text{RF}}, \mathbf{w}_{\text{BB}} }{\text{max}} \;\; \underset{k \in \mathcal{K}}{\text{min}} \quad \frac{ |\bar{\mathbf{h}}^H_{k} \mathbf{F}_{\text{RF}} \mathbf{w}_{\text{BB}}|^2}{\|\mathbf{F}_{\text{RF}} \mathbf{w}_{\text{BB}}\|^2} \nonumber \\
&& \;\; \text{s.t.} \quad  \mathbf{F}_{\text{RF}} \in \mathcal{C}_{\text{set}},
\end{eqnarray}
where we unify the power constraint into the objective function. For certain $\mathbf{F}_{\text{RF}}$, the optimum value (denoted by $t$) of the objective function in problem \eqref{eq:hybridpre2} is upper bounded, i.e.,
\begin{eqnarray} \label{eq:obser}
t &=& \underset{k \in \mathcal{K}}{\text{min}} \; \bigg\{ \frac{\mathbf{w}^H_{\text{BB}} \mathbf{F}^H_{\text{RF}} \bar{\mathbf{h}}_{k} \bar{\mathbf{h}}^H_{k} \mathbf{F}_{\text{RF}} \mathbf{w}_{\text{BB}}}{\mathbf{w}^H_{\text{BB}} \mathbf{F}^H_{\text{RF}} \mathbf{F}_{\text{RF}} \mathbf{w}_{\text{BB}}} \bigg\} \\ \label{obser2}
&\le& \underset{k \in \mathcal{K}}{\text{min}} \; \Big\{ \lambda_{\text{max}} \Big( (\mathbf{F}^H_{\text{RF}} \mathbf{F}_{\text{RF}})^{-1} \mathbf{F}^H_{\text{RF}}\bar{\mathbf{h}}_{k} \bar{\mathbf{h}}^H_{k} \mathbf{F}_{\text{RF}} \Big)\Big\} \\ \label{obser3}
&=& \underset{k \in \mathcal{K}}{\text{min}} \; \big\{ \bar{\mathbf{h}}^H_{k} \mathbf{F}_{\text{RF}} (\mathbf{F}^H_{\text{RF}} \mathbf{F}_{\text{RF}})^{-1} \mathbf{F}^H_{\text{RF}}\bar{\mathbf{h}}_{k} \big\},
\end{eqnarray}
where \eqref{obser2} is obtained from the generalized eigenvalue of $\mathbf{F}^H_{\text{RF}} \bar{\mathbf{h}}_{k} \bar{\mathbf{h}}^H_{k} \mathbf{F}_{\text{RF}} $ and $\mathbf{F}^H_{\text{RF}} \mathbf{F}_{\text{RF}}$. We assume $\mathcal{C}$ with full column rank and then $\mathbf{F}^H_{\text{RF}} \mathbf{F}_{\text{RF}}$ is invertible.

A selection of $\mathbf{F}_{\text{RF}} \in \mathcal{C}_{\text{set}}$, that leads to a small upper bound, constrains the optimum $t$ to a small value and hence is unlikely to be the optimal RF precoder. In this light, we propose to search $\mathbf{F}_{\text{RF}}$ in descending order of \eqref{obser3} and only select a limited number of $\mathbf{F}_{\text{RF}}$ that obtains the largest upper bound values. The low complexity search algorithm is outlined in Algorithm 1.

\begin{table}[t] \label{alg1}
\caption*{\hspace{-35pt}Algorithm 1: Precoding Design}
\small
\begin{tabular}{ll}
\hline
1: \textbf{Pre-processing}: Select a subset of $\mathbf{F}_{\text{RF}} \in \mathcal{C}_{\text{set}}$ with the\\
\hspace{70pt} largest $I$ upper bound values in \eqref{obser3} \\
2: \textbf{For} $i \in [1, I]$\\
3: \quad Compute the effective channel $\hat{\mathbf{h}}^{(i)}_k, \forall k$ with $\mathbf{F}^{(i)}_{\text{RF}}$ \\
4: \quad Apply any algorithm stated in section \ref{SDR} to find \\ \quad \;\;\; the optimum $t^{(i)}$ and the corresponding $\mathbf{w}^{(i)}_{\text{BB}}$ \\
5: \textbf{End}\\
6: The $\mathbf{F}_{\text{RF}}$ and $\mathbf{w}_{\text{BB}}$ that obtains the largest $t$ are chosen \\  \;\;\; as the solution of \eqref{eq:hybridpre2} \\
\hline
\end{tabular}
\end{table}

\emph{Remark 1}: The complexity of Algorithm 1 is approximated by {\small$I \cdot C_d$}, where {\small$C_d$} indicates the complexity of algorithm applied in step 4. We note that {\small$I \in [1, I_{\text{exs}}]$}. The optimal $\mathbf{F}_{\text{RF}}$ can be obtained by exhaustive search ({\small $I = I_{\text{exs}}$}) while the proposed RF precoder design can achieve nearly the same max-min fairness with {\small$I \ll I_{\text{exs}}$}. The corresponding complexity reduction factor can be approximated by {\small$I_{\text{exs}}/I$}.

\emph{Remark 2}: The RF codebook design acts as one of the determinant factors for the max-min performance. Though the optimal RF codebook is still unknown, good candidates can be recognized by exploiting second-order channel statistics or array structure. For example, we can normalize each entry of the dominant eigenvectors of channel covariance matrices and collect them as the RF codebook. Otherwise, we can design $\mathcal{C}$ as the steering vectors with uniformly distributed AoDs. Moreover, when we take implementation complexity into account, a DFT-based codebook can be employed where any RF precoder can be implemented by a static (DFT) phase shifting network together with a RF switch.

\subsection{Achieving Optimum with a Reduced Number of RF Chains} \label{mmwchannel}
The proposed algorithm in section \ref{algorithm} is applicable to arbitrary channels. By imposing a predefined structure on the RF precoder design, the hybrid precoding can only provide a suboptimal performance compared with the fully digital precoding case. In this section, we consider a physical finite scattering channel model that has been investigated for Massive/Mm-wave MIMO systems \cite{Omar2014,Ngo2013}. The channel vector from user $k$ is defined as
\begin{eqnarray} \label{eq:channel}
\mathbf{h}_k = \sqrt{\frac{M}{L_k}} \sum^{L_k}_{l=1} g^l_{k} \, \mathbf{a}(\phi^l_k,\theta^l_k) =  \sqrt{\frac{M}{L_k}} \mathbf{A}_k \mathbf{g}_k,
\end{eqnarray}
where {\small$\mathbf{A}_k = [\mathbf{a}(\phi^1_k,\theta^1_k),\cdots, \mathbf{a}(\phi^{L_k}_k,\theta^{L_k}_k)] \in \mathbb{C}^{M \times L_k}$} contains $L_k$ steering vectors and $\mathbf{g}_{k}$ is the path gain vector. The factor {\small $\sqrt{M/L_k}$} is used to normalize the channel, i.e., $\mathbb{E}(\|\mathbf{h}_k\|^2) = M$. Under the plane wave and balanced narrowband array assumptions, the array steering vector can be written as \cite[Ch.~2]{Bruno2013}
\begin{eqnarray} \label{eq:steer}
\mathbf{a}(\phi^l_k,\theta^l_k) = \frac{1}{\sqrt{M}} \big[e^{-j f_1(\phi^l_k,\theta^l_k)},\cdots, e^{-j f_{M}(\phi^l_k,\theta^l_k)}\big]^T,
\end{eqnarray}
where $f_{m}(\phi,\theta)$ is a function of the azimuth ($\phi$) and elevation ($\theta$) angle-of-departure (AoD). Consider first the single-path channels (i.e., $L_k = 1$) and denote a collection of the steering vector of each user by a $M \times K$ matrix {\small $\mathbf{V}$} (e.g., {\small $\mathbf{V}=[\mathbf{a}(\phi_1,\theta_1),\cdots, \mathbf{a}(\phi_K,\theta_K)]$}). The following theorem characterizes the max-min performance of multicasting driven by a limited number of RF chain.

$\textbf{Theorem 1:}$ When the BS has a prior knowledge of AoDs\footnote{Subspace methods (e.g., root-MUSIC and ESPRIT) and compressed sensing algorithms (e.g., OMP and BP) can be applied to identify the distinct path arrivals \cite{Berger2010}.} and all channels are single-path, the hybrid precoding with $\mathbf{F}_{\text{RF}} = \mathbf{V}$ driven by only $K$ RF chains achieves the same max-min fairness as with fully digital precoding driven by a sufficient number of RF chains (i.e., $N=M$).
\begin{IEEEproof}
We denote the (tall) channel matrix by $\mathbf{H} = [\mathbf{h}_1,\cdots, \mathbf{h}_K]$. Focusing on the fully digital precoder design in problem \eqref{eq:fullydig}, we note that $\forall \, \mathbf{w} \in \text{span}^{\bot}(\mathbf{H})$ makes the value of the objective function zero. Hence, the optimal solution $\mathbf{w^*} \in \text{span}(\mathbf{H})$ is written as a linear combination of $\mathbf{h}_k$, i.e.,
\begin{eqnarray} \label{eq:linear}
\mathbf{w^*} = \sum_{k=1}^K b_k \mathbf{h}_k = \sum_{k=1}^K c_k \mathbf{a}(\phi_k,\theta_k) = \mathbf{F}_{\text{RF}} \mathbf{w}_{\text{BB}},
\end{eqnarray}
where $b_k$ denote the coefficients and $\mathbf{w}_{\text{BB}} = [c_1, \cdots, c_K]^T$ with $c_k = \sqrt{M} b_k g_k$. Since the steering vector collection matrix $\mathbf{V}$ satisfies inherently the constant modulus constraint, \eqref{eq:linear} implies that the optimal fully digital precoder can be equivalently implemented by a hybrid precoding structure without loss of performance.
\end{IEEEproof}

By reusing \eqref{eq:linear}, we are able to generalize Theorem 1 into multi-path channels (i.e., $L_k \ge 1$) and thereby present the following Corollary without proof.

$\textbf{Corollary 1:}$ When the BS has a prior knowledge of AoDs and at least $N$ RF chains with $N = \sum_k L_k \le M$, and let $\mathbf{V} \in \mathbb{C}^{M \times N}$ collect the steering vectors of all users, the hybrid precoder with $\mathbf{F}_{\text{RF}} = \mathbf{V}$ achieves the same max-min fairness as with fully digital precoder.


\section{SIMULATION RESULTS} \label{numresults}

\begin{figure*}[t]
  \centering
\subfigure[M = 6, K = 2]{\label{mult1} \includegraphics[width = 0.34\textwidth]{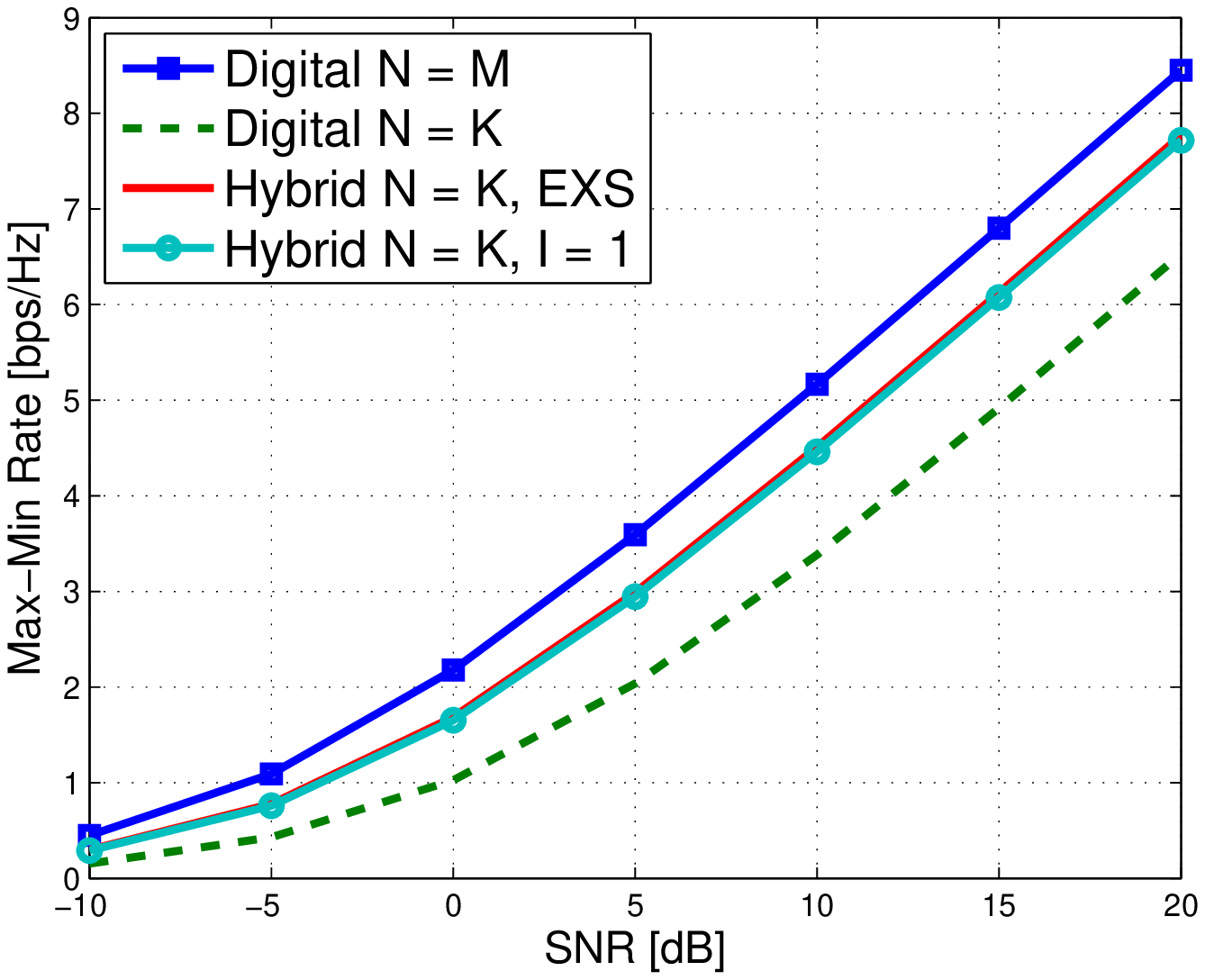}} \hspace{-18pt}
\subfigure[M = 8, K = 3]{\label{mult2} \includegraphics[width = 0.34\textwidth]{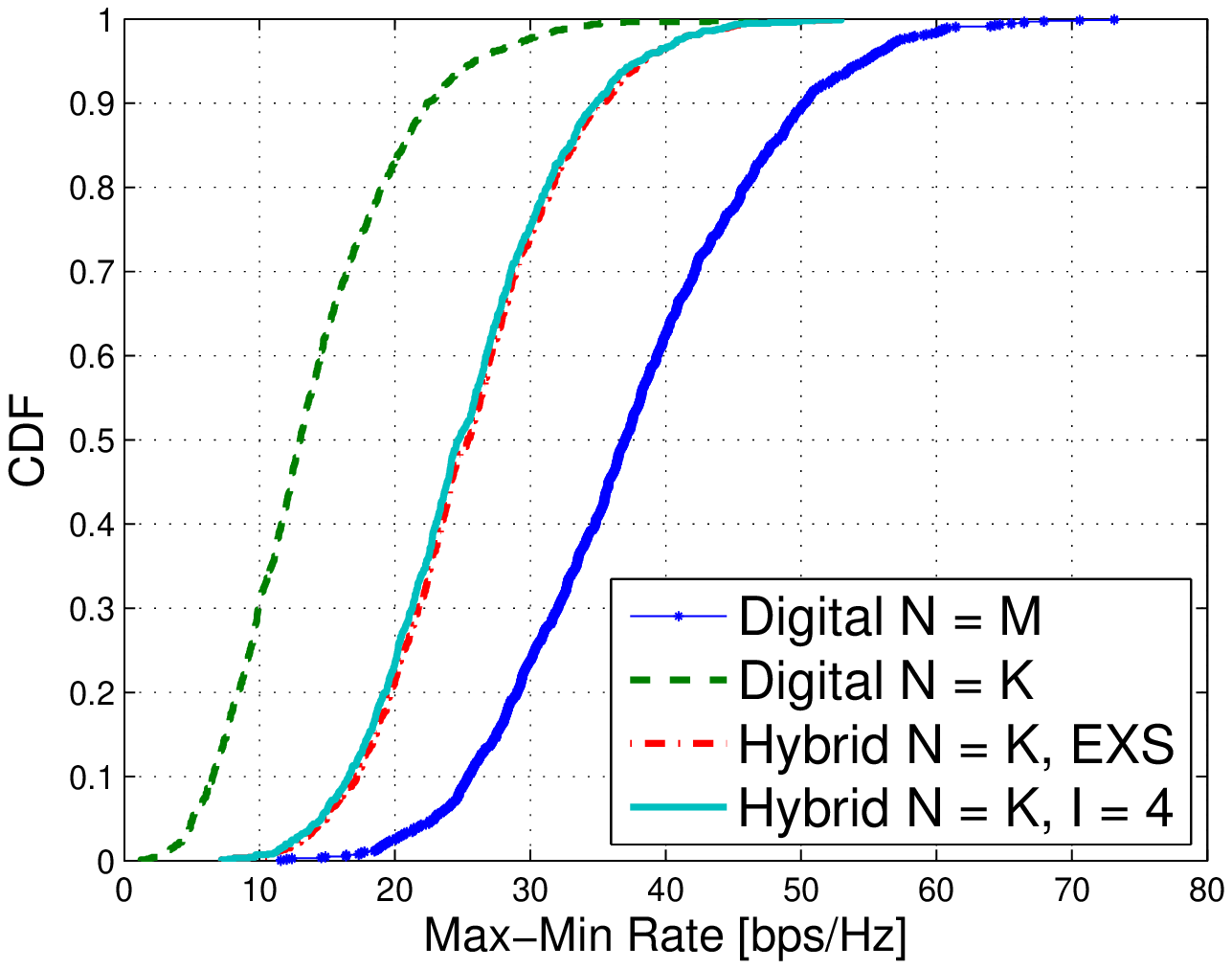}} \hspace{-18pt}
\subfigure[M = 10,K = 3]{\label{mult3} \includegraphics[width = 0.34\textwidth]{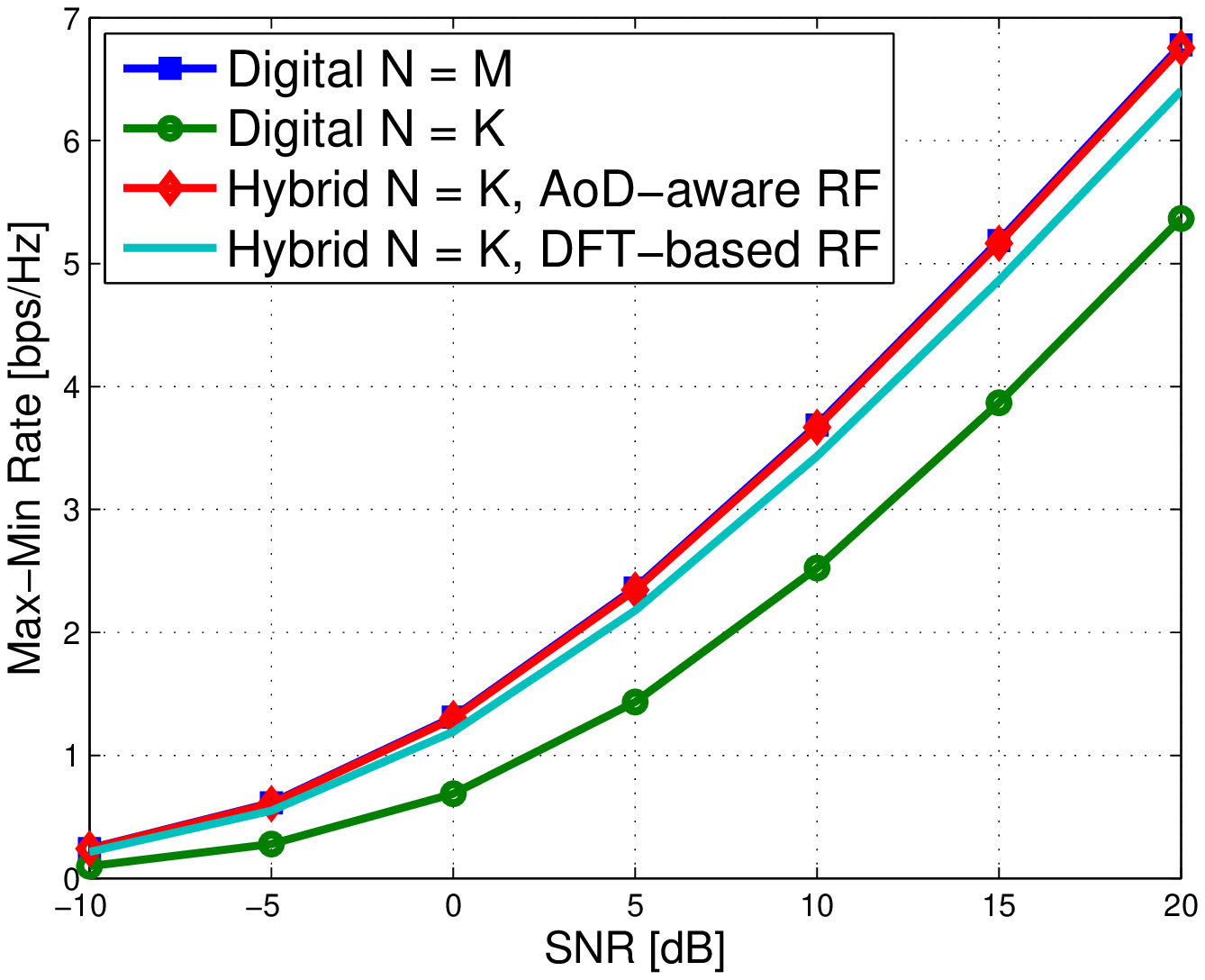}}
\caption{max-min rate of the proposed hybrid precoding design. } \label{mult}
\end{figure*}

In this section, we numerically compare the proposed low complexity algorithm with several baselines: Baseline 1 assumes a sufficient number of RF chains and exploits fully digital precoding (i.e., $M = N$, denoted by `Digital N = M' in Fig. \ref{mult}). Baseline 2 assumes a random antenna subset selection and performs digital precoding over that subset (i.e., $M' = N = K$, denoted by `Digital N = K' in Fig. \ref{mult}). Intuitively, baselines $1\&2$ place a upper and lower bound on the performance of hybrid precoder, respectively. Baseline 3 assumes a limited number of RF chains and exploits hybrid precoding with exhaustive search for RF precoder (i.e., $M > N = K$, denoted by `Hybrid N = K, EXS' in Fig. \ref{mult}). DFT-based RF codebook is taken as an example. Since the interest of this letter is the RF precoder design, we compute all the digital precoders by simply using SDR. A Gaussian randomization procedure is carried out to obtain a feasible solution. Moreover, the max-min rate is investigated by averaging over 1000 random channel realizations.

In Fig. \ref{mult1} and \ref{mult2}, we evaluate the max-min rate performance under independent and identical distributed (i.i.d.) Rayleigh fading channels. Consider a system with $M = 6, K = 2$, Fig. \ref{mult1} shows that the hybrid precoding structure achieves a max-min rate gain over baseline 2 (e.g., 1 bps/Hz beyond 5 dB SNR). It implies that a cost-efficient and well-designed phase shifter network can reap benefits. Compared with baseline 3 {\small $\left(I_{\text{exs}} = \binom{N}{M} = 15 \right)$}, the proposed algorithm with $I = 1$ highly reduces the search complexity while keeping the same performance. In Fig. \ref{mult2}, we consider $M = 8, K = 3,$ SNR = 10 dB and plot the cumulative density function (CDF) of the max-min rate of each approach. Likewise, the proposed low complexity algorithm with $I = 4$ achieves similar performance as with the exhaustive search ($I_{\text{exs}} = 56$). Moreover, we observe that the hybrid precoding with RF precoder selected from a predefined codebook is outperformed by the fully digital precoding. This performance loss is mainly due to the mismatch between the constant magnitude constrained RF precoder and the non-constant magnitude (non-sparse) i.i.d. Rayleigh channels.

Fig. \ref{mult3} examines the max-min rate in single-path channels. Consider a uniform linear array (ULA) with $M = 10$ isotropic antennas serving $K = 3$ users. The steering vector is given by {\small $\mathbf{a}(\phi^l_k) = \frac{1}{\sqrt{M}} [1, e^{-j 2 \pi\frac{D}{\lambda} \cos(\phi^l_k)},\cdots, e^{-j 2 \pi\frac{(M-1)D}{\lambda} \cos(\phi^l_k)}]^T$}, where {\small$D = \lambda/2$} is the half-wavelength antenna spacing and $\phi^l_k$ is uniformly distributed between $0$ and $2\pi$. Fig. \ref{mult3} shows that the AoD-aware RF precoder design (Theorem 1) outperforms the low complexity DFT-based approach (Algorithm 1 with $I = 1$). The former dynamic design requires a prior knowledge of AoDs while the latter less-flexible design facilitates ease of implementation. In addition, Theorem 1 is validated, namely, the hybrid precoder with a limited number of RF chains can achieve the same max-min rate as with fully digital precoder. It implies that channel sparsity can be exploited to reduce the number of costly RF chains without a compromise on the performance. The intuition behind Theorem 1 and Corollary 1 is explained as follows. Under a limited scattering model, the $M \times 1$ channel vector is characterized by a few AoDs ($\psi^{l}$). The max-min fairness enabled by a high-dimensional ($M$) digital precoder can be obtained by a $M \times N$ transformation matrix (i.e., the RF precoder) that accurately captures the channel gain and a low-dimensional ($N \ge K$) digital precoder that takes care of the fairness amongst $K$ users.

Moreover, when the channels have multi-path of $\sum_k L_k \ge M$, Theorem 1 and Corollary 1 are not achievable. In this case, suppose the BS has the knowledge of AoDs, the low complexity algorithm in section \ref{algorithm} can be applied to select the RF precoder from the codebook that collects the steering vectors of all users. The basic idea is to exclude those selections that constrain the upper bound in \eqref{obser3} and to search among the rest. This codebook design would perform better than various codebooks discussed in remark 2, since it mostly captures the directions of the users' channels.

\section{CONCLUSION} \label{conclusion}
In this letter, we have investigated a hybrid precoding method for multicasting with a limited number of RF chains. We have proposed a low complexity search algorithm to determine the RF precoder and validated its near optimality in terms of max-min rate. For limited scattering channels, we proved a simple condition under which the hybrid precoding incurs no loss of optimality with respect to the fully digital precoding case. Moreover, the hybrid precoding in multi-group/cell multicasting is intriguing and is left for future work.

\bibliographystyle{IEEEtran}
\bibliography{reference}

\end{document}